\documentclass[12pt,english]{amsart}
  \usepackage[latin9]{inputenc}
\usepackage{geometry}
\usepackage{amsthm}
\usepackage{graphicx}
\usepackage{amssymb}

\numberwithin{equation}{section} 
\numberwithin{figure}{section} 
\theoremstyle{plain}
\theoremstyle{plain}
\newtheorem{thm}{Theorem}
  \theoremstyle{plain}
  \newtheorem{prop}[thm]{Proposition}
  \theoremstyle{definition}
  \newtheorem{defn}[thm]{Definition}
  \theoremstyle{plain}
  \newtheorem{lem}[thm]{Lemma}
  \theoremstyle{remark}
  \newtheorem{rem}[thm]{Remark}
  \theoremstyle{plain}
  \newtheorem{cor}[thm]{Corollary}
  \theoremstyle{plain}
  \newtheorem{conjecture}[thm]{Conjecture}

\newcommand{\nwc}{\newcommand}
\nwc{\nwt}{\newtheorem}
\nwt{coro}{Corollary}

\nwc{\mf}{\mathbf} 
\nwc{\blds}{\boldsymbol} 
\nwc{\ml}{\mathcal} 

\nwc{\lam}{\lambda}
\nwc{\del}{\delta}
\nwc{\Del}{\Delta}
\nwc{\Lam}{\Lambda}
\nwc{\elll}{\ell}

\nwc{\IA}{\mathbb{A}} 
\nwc{\IB}{\mathbb{B}} 
\nwc{\IC}{\mathbb{C}} 
\nwc{\ID}{\mathbb{D}} 
\nwc{\IE}{\mathbb{E}} 
\nwc{\IF}{\mathbb{F}} 
\nwc{\IG}{\mathbb{G}} 
\nwc{\IH}{\mathbb{H}} 
\nwc{\IN}{\mathbb{N}} 
\nwc{\IP}{\mathbb{P}} 
\nwc{\IQ}{\mathbb{Q}} 
\nwc{\IR}{\mathbb{R}} 
\nwc{\IS}{\mathbb{S}} 
\nwc{\IT}{\mathbb{T}} 
\nwc{\IZ}{\mathbb{Z}} 
\def\bbbone{{\mathchoice {1\mskip-4mu {\rm{l}}} {1\mskip-4mu {\rm{l}}}
{ 1\mskip-4.5mu {\rm{l}}} { 1\mskip-5mu {\rm{l}}}}}

\nwc{\va}{{\bf a}}
\nwc{\vb}{{\bf b}}
\nwc{\vc}{{\bf c}}
\nwc{\vd}{{\bf d}}
\nwc{\ve}{{\bf e}}
\nwc{\vf}{{\bf f}}
\nwc{\vg}{{\bf g}}
\nwc{\vh}{{\bf h}}
\nwc{\vi}{{\bf i}}
\nwc{\vj}{{\bf j}}
\nwc{\vk}{{\bf k}}
\nwc{\vl}{{\bf l}}
\nwc{\vm}{{\bf m}}
\nwc{\vn}{{\bf n}}
\nwc{\vo}{{\it o}}
\nwc{\vp}{{\bf p}}
\nwc{\vq}{{\bf q}}
\nwc{\vr}{{\bf r}}
\nwc{\vs}{{\bf s}}
\nwc{\vt}{{\bf t}}
\nwc{\vu}{{\bf u}}
\nwc{\vv}{{\bf v}}
\nwc{\vw}{{\bf w}}
\nwc{\vx}{{\bf x}}
\nwc{\vy}{{\bf y}}
\nwc{\vz}{{\bf z}}
\nwc{\bal}{\boldsymbol{\alpha}}
\nwc{\bep}{\boldsymbol{\epsilon}}
\nwc{\bnu}{\boldsymbol{\nu}}
\nwc{\bmu}{\boldsymbol{\mu}}

\nwc{\balpha}{\blds{\alpha}}
\nwc{\bk}{\blds{k}}

\nwc{\bm}{\blds{m}}
\nwc{\bp}{\blds{p}}
\nwc{\bq}{\blds{q}}
\nwc{\bn}{\blds{n}}
\nwc{\bu}{\blds{u}}
\nwc{\bv}{\blds{v}}
\nwc{\bw}{\blds{w}}
\nwc{\bx}{\blds{x}}
\nwc{\bxi}{\blds{\xi}}
\nwc{\by}{\blds{y}}
\nwc{\bz}{\blds{z}}

\nwc{\cA}{\ml{A}}
\nwc{\cB}{\ml{B}}
\nwc{\cC}{\ml{C}}
\nwc{\cD}{\ml{D}}
\nwc{\cE}{\ml{E}}
\nwc{\cF}{\ml{F}}
\nwc{\cG}{\ml{G}}
\nwc{\cH}{\ml{H}}
\nwc{\cI}{\ml{I}}
\nwc{\cJ}{\ml{J}}
\nwc{\cK}{\ml{K}}
\nwc{\cL}{\ml{L}}
\nwc{\cM}{\ml{M}}
\nwc{\cN}{\ml{N}}
\nwc{\cO}{\ml{O}}
\nwc{\cP}{\ml{P}}
\nwc{\cQ}{\ml{Q}}
\nwc{\cR}{\ml{R}}
\nwc{\cS}{\ml{S}}
\nwc{\cT}{\ml{T}}
\nwc{\cU}{\ml{U}}
\nwc{\cV}{\ml{V}}
\nwc{\cW}{\ml{W}}
\nwc{\cX}{\ml{X}}
\nwc{\cY}{\ml{Y}}
\nwc{\cZ}{\ml{Z}}

\nwc{\tA}{\widetilde{A}}
\nwc{\tB}{\widetilde{B}}

\nwc{\To}{\longrightarrow} 

\nwc{\ad}{\operatorname{ad}}
\nwc{\eps}{\epsilon}
\nwc{\ep}{\epsilon}
\nwc{\vareps}{\varepsilon}

\def\ep{\epsilon}

\def\sq2{\sqrt{2}}

\def\defeq{\stackrel{\rm def}{=}}
\def\t2{{\mathbb T}^2}
\def\tt2{{\mathbb T}^2}
\def\s2{{\mathbb S}^2}

\nwc{\diag}{\operatorname{diag}}
\nwc{\rest}{\restriction}
\nwc{\diam}{\operatorname{diam}}
\nwc{\Res}{\operatorname{Res}}
\nwc{\Spec}{\operatorname{Spec}}
\nwc{\Vol}{\operatorname{Vol}}
\nwc{\Op}{\operatorname{Op}}
\nwc{\Oph}{\operatorname{Op}_\hbar}
\nwc{\Prob}{\operatorname{Prob}}
\nwc{\supp}{\operatorname{supp}}
\nwc{\rank}{\operatorname{rank}}
\nwc{\esssup}{\operatorname{ess-sup}}
\nwc{\essinf}{\operatorname{ess-inf}}

\nwc{\Span}{\operatorname{span}}

\def\hto0{\xrightarrow{h\to 0}}

\def\rto0{\xrightarrow{r\to 0}}

\nwc{\la}{\langle}
\nwc{\ra}{\rangle}
\nwc{\lp}{\left(}
\nwc{\rp}{\right)}
\renewcommand{\Re}{\operatorname{Re}}
\renewcommand{\Im}{\operatorname{Im}}

\usepackage{babel}

\begin{document}

\title{Spectral theory of damped quantum chaotic systems}

\author{Stéphane Nonnenmacher}

\thanks{This work has been partially
    supported by the grant ANR-09-JCJC-0099-01 of the Agence Nationale
  de la Recherche.}

\address{Institut de Physique théorique, CEA-Saclay, unité de recherche associée
au CNRS, 91191 Gif-sur-Yvette, France}

\begin{abstract}
We investigate the spectral distribution of the damped wave equation
on a compact Riemannian manifold, especially in the case of a metric
of negative curvature, for which the geodesic flow is Anosov. The
main application is to obtain conditions (in terms of the geodesic
flow on $X$ and the damping function) for which the energy of the
waves decays exponentially fast, at least for smooth enough initial
data. We review various estimates for the high frequency spectrum
in terms of dynamically defined quantities, like the value distribution
of the time-averaged damping. We also present a new condition for
a spectral gap, depending on the set of minimally damped trajectories.
\end{abstract}
\maketitle

\section{Introduction}

\subsection{Spectrum of the damped wave equation\label{sub:intro}}

Given a Riemannian manifold $(X,g)$ and a \emph{damping function}
$a\in C^{\infty}(X,\IR_{+})$, we are interested in the solutions
of the damped wave equation\emph{ }(DWE) \begin{equation}
(\partial_{t}^{2}-\Delta+2a(x)\partial_{t})v(x,t)=0,\quad v(x,0)=v_{0},\;\partial_{t}v(x,0)=v_{1}\,.\label{eq:DWE}\end{equation}
The natural setting is to take intial data $(v_{0},v_{1})$ in the
space $\cH\defeq H^{1}(X)\times L^{2}(X)$. This equation is equivalent
with the system \begin{equation}
\left(i\partial_{t}+\cA\right)\bv(t)=0,\quad\cA\defeq\left(\begin{array}{cc}
0 & I\\
-\Delta & -2ia\end{array}\right),\quad\bv(0)\in\cH,\label{eq:DWE-A}\end{equation}
with the correspondence $\bv(t)=\left(v(t),i\partial v(t)\right)$.
$\cA$ generates a strongly continuous semigroup on $\cH$, so the
solution to (\ref{eq:DWE},\ref{eq:DWE-A}) reads\begin{equation}
\bv(t)=e^{-it\cA}\bv(0).\label{eq:semigroup}\end{equation}
We will always assume that the damping is nontrivial, $a\not\equiv0$.
Apart from the constant solution $\bv(t)=(1,0)$, all solutions then
decay. Physically, this is expressed by the fact that the \emph{energy}
of the waves, \begin{equation}
E(v(t))=\frac{1}{2}\left(\left\Vert \nabla v(t)\right\Vert ^{2}+\left\Vert \partial_{t}v(t)\right\Vert ^{2}\right),\label{eq:energy}\end{equation}
will decay to zero when $t\to\infty$ for any $\bv(0)\in\cH$. In
some sense, the waves are stabilized by the damping.

To analyze this decay, it it natural to try to expand the solution
in terms of the spectrum of $\cA.$ This spectrum is discrete, consisting
of countably many complex eigenvalues $\left\{ \tau_{n}\right\} $
with $\Re\tau_{n}\to\pm\infty$. It can be obtained by solving the
generalized eigenvalue equation\begin{equation}
\left(-\Delta-\tau^{2}-2ia\tau\right)u=0.\label{eq:eigenvalue}\end{equation}
The following properties are easily shown ($a_{\min}\defeq\min_{x\in X}a(x)$,
similarly for $a_{\max}$). 
\begin{prop}
\cite{Leb93}\label{lem:strip-a}All eigenvalues except $\tau_{0}=0$
satisfy $\Im\tau_{n}<0$. 

If $\Re\tau_{n}\neq0$ then $-\Im\tau_{n}\in[a_{\min},a_{\max}]$. 

The spectrum is symmetric w.r.to the imaginary axis.
\end{prop}
To each eigenvalue $\tau_{n}$ corresponds a quasi-stationary mode
$u_{n}(x)$, an eigenstate $\bu_{n}=(u_{n},\tau_{n}u_{n})$ of $\cA$
reads, and a solution $v_{n}(t,x)=e^{-it\tau_{n}}u_{n}(x)$ of the
DWE. Hence $\Im\tau_{n}$ represents the \emph{quantum decay rate}
of the mode $u_{n}$. 

If $a_{\min}>0$, then the energy decays exponentially, uniformly
for initial data $\bv(0)\in\cH$. Precisely, there exists $\gamma>0$,
$C>0$ such that\begin{equation}
E(\bv(t))\leq C\, e^{-2\gamma t}\, E(\bv(0)),\qquad\forall\bv(0)\in\cH,\quad\forall t\geq0.\label{eq:decay-GCC}\end{equation}

\subsection{The Geometric Control Condition}

The condition $a_{\min}>0$ is not necessary to ensure such a uniform
exponential decay. Since all eigenvalues except $\tau_{0}=0$ satisfy
$\Im\tau_{n}<0$, for any $C>0$ the subspace $\cH_{C}\subset\cH$
spanned by the eigenstates $\left\{ \bu_{n},\:\left|\Re\tau_{n}\right|\leq C\right\} $
is finite dimensional, and for any initial data $\bv(0)\subset\cH_{C}$
the energy will decay exponentially. Hence, the failure of exponential
decay can only come from the behaviour of waves at high frequency.
In this high frequency limit, a natural connection can be made with
the classical ray dynamics on $X$, equivalently the geodesic flow
$\Phi^{t}$ on the unit cotangent bundle $S^{*}X$ (see $\S$\ref{sub:High-frequency}).
Using this connection, Rauch and Taylor \cite{RauTay75} showed that
the uniform exponential decay (\ref{eq:decay-GCC}) is equivalent
with the Geometric Control Condition (GCC), which states that every
geodesic meets the damping region $\left\{ x\in X,\, a(x)>0\right\} $
(due to the compactness of $S^{*}X$ each geodesic does it within
some time $T_{0}>0$). 

This condition can be expressed in terms of the \emph{time averages}
of the damping, namely the functions\begin{equation}
\la a\ra_{t}(\rho)=\frac{1}{t}\int_{0}^{t}a\circ\Phi^{s}(\rho)\, ds,\quad\rho\in T^{*}X,\; t>0.\label{eq:a_T}\end{equation}
GCC is equivalent to the fact that, for $t>0$ large enough (say,
$t>2T_{0}$), the function $\la a\ra_{t}$ is strictly positive on
$S^{*}X$. %
\begin{figure}
\includegraphics[scale=0.4]{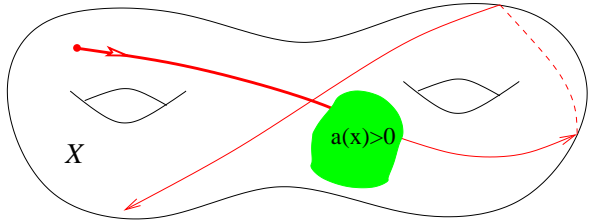}\caption{A damped geodesic.}
\end{figure}
Lebeau \cite{Leb93} generalized this result to the case of manifolds
with boundaries. He also showed that the optimal decay rate $\gamma$
is given by $\min\left(G,a_{-}\right)$, where $G=\inf\left\{ -\Im\tau_{n},\:\tau_{n}\neq0\right\} $
is the spectral gap, while \begin{equation}
a_{-}\defeq\lim_{t\to\infty}\min_{S^{*}X}\la a\ra_{t}\label{eq:a_-}\end{equation}
is the minimal asymptotic damping. We will explain the relevance of
$a_{-}$ in $\S$\ref{sub:QDR}. 

Koch and Tataru \cite{KoTa94} studied the same question in a more
general context (case of manifolds with boundaries, and of a damping
taking place both in the {}``bulk'' and on the boundary). They showed
that the averages $\la a\ra_{t}$ govern the decay of the semigroup,
up to a compact subspace. 
\begin{thm}
\cite[Thm 2]{KoTa94}For each $\eps>0$ and each $t>0$ there exists
a subspace $\cH_{\eps,t}\subset\cH$ of finite codimension such that
\[
\left\Vert e^{-it\cA}\right\Vert _{\cH_{\eps,t}\to\cH}\leq\exp\left\{ -t\min_{S^{*}X}\la a\ra_{t}\right\} +\eps.\]
For any subspace $\cH_{1}\subset\cH$ of finite codimension, \[
\exp\left\{ -t\min_{S^{*}X}\la a\ra_{t}\right\} \leq\left\Vert e^{-it\cA}\right\Vert _{\cH_{1}\to\cH}.\]

\end{thm}
A consequence of this result is the characterization of the \emph{Fredholm
spectrum} of the semigroup. In the present situation of damped waves
on a compact $X$ withough boundary, their result states%
\footnote{under the condition that, for any $T>0$, the set of $T$-periodic
geodesics has measure zero. %
} that this Fredholm spectrum is given by the annulus $\left\{ z\in\IC,\: e^{-ta_{+}}\leq|z|\leq e^{-ta_{-}}\right\} $. 

Notice that it is possible to have a positive gap $G>0$ and at the
same time $a_{-}=0$ (failure of the GCC), see e.g. an example in
\cite{Ren94}. This reflects the fact that the spectrum of the semigroup
$e^{-it\cA}$ is not controlled by the spectrum of generator $\cA$,
a frequent problem for \emph{nonnormal} generators like $\cA$.

\subsection{Beyond Geometric control: a few cases of uniform energy decay for
regular data}

If GCC fails, one can construct initial data $\bv(0)\in\cH$ such
that $E(\bv(t))$ decays arbitrarily slowly \cite{Leb93}. Yet, it
is possible to show that more regular data, say in some Sobolev space
$\cH^{s}=H^{s+1}\times H^{s}$, $s>0$, decay in some uniform way.
In the most general situation, Lebeau showed a logarithmic decay of
the energy \cite[Thm 1]{Leb93},\[
E(\bv(t))^{1/2}\leq C_{s}\,\frac{\log\left[3+\log(3+t)\right]}{\log(3+t)^{s}}\left\Vert \bv\right\Vert _{\cH^{s}},\quad\forall\bv=\bv(0)\in\cH^{s},\:\forall t\geq0.\]
The proof uses Carleman estimates, which imply some control of the
resolvent $\left(\tau-\cA\right)^{-1}$ in a strip $\left\{ \Re\tau\geq C,\;|\Im\tau|\leq f(\Re\tau)\right\} $
for some function $f(r)$ which decays exponentially as $r\to\infty$.
By interpolation, proving such an estimate for some $s_{0}>0$ implies
a similar estimate for all $s>0$. 

In specific dynamical situations, one can show resolvent estimates
in a larger strip, e.g. for some function $f(r)$ decaying algebraically:
one then gets an algebraic decay $\cO(t^{-\gamma_{s}}\left\Vert \bv\right\Vert _{\cH^{s}})$
for the energy of regular data. Burq and Hitrik showed that this is
the case for damped waves in the (chaotic) \emph{stadium billiard},
if the damping function vanishes only in some part of the central
rectangle, so that the \emph{set of undamped trajectories},\begin{equation}
K\defeq\left\{ \rho\in S^{*}X,\:\forall t\in\IR,\; a(\Phi^{t}(\rho))=0\right\} ,\label{eq:undamped-set}\end{equation}
consists in a collection of (neutrally stable) {}``bouncing ball''
trajectories \cite{BuHi07}. 

Christianson studied the case where the set $K$ consists in a single
\emph{hyperbolic} closed geodesic: the energy decay is bounded by
a stretched exponential $\cO(e^{-\gamma_{s}t^{1/2}}\left\Vert \bv\right\Vert _{\cH^{s}})$
\cite[Thm 5']{Chris07}\cite{Chris11}. The hyperbolicity of the geodesic
induces a strong dispersion of the waves, responsible for this fast
decay. 

In $\S$\ref{sub:pressure-criterium} and \ref{sub:thickness-condition}
we will present situations, with $(X,g)$ a manifold of negative curvature,
for which the energy of regular data decays exponentially: \begin{equation}
E(\bv(t))^{1/2}\leq C_{s}\, e^{-\gamma_{s}t}\,\left\Vert \bv\right\Vert _{\cH^{s}},\quad\forall\bv\in\cH^{s},\:\forall t\geq0.\label{eq:decay-loss}\end{equation}
The proof proceeds by controlling the resolvent $\left(\tau-\cA\right)^{-1}$
in a strip $\left\{ \Re\tau>C,\:|\Im\tau|\leq c\right\} $, and then
uses standard arguments \cite{Chris09}. Here as well, the resolvent
bounds are based on certain \emph{hyperbolic dispersion} estimates.
See the Corollaries \ref{cor:pressure-DWE} and \ref{cor:thickness-resolvent}
for the precise dynamical conditions, which involve the interplay
between the flow and the damping function.

\subsection{High frequency limit -- semiclassical formulation and generalization\label{sub:High-frequency}}

As explained above, the decay of the energy is mainly governed by
the high frequencies $\Re\tau\gg1$. A semiclassical formulation of
the problem was used in \cite{Sjo00}, allowing to use $\hbar$-pseudodifferential
techniques. Take an effective Planck's {}``constant'' $0<\hbar\ll1$.
In order to study the frequencies $\Re\tau=\hbar^{-1}+\cO(1)$, we
perform the rescaling \begin{equation}
\tau=\frac{\sqrt{2z}}{\hbar}\quad\mbox{with }z\:\mbox{in the disk }D(1/2,C\hbar).\label{eq:tau-z}\end{equation}
The equation (\ref{eq:eigenvalue}) becomes\begin{equation}
(P(\hbar,z)-z)u=0,\qquad P(\hbar,z)=-\frac{\hbar^{2}\Delta}{2}-i\hbar\sqrt{2z}\, a=-\frac{\hbar^{2}\Delta}{2}-i\hbar a+\cO(\hbar^{2}).\label{eq:semiclass}\end{equation}
More generally, we consider operators of the type \begin{equation}
P(\hbar,z)=-\frac{\hbar^{2}\Delta}{2}+i\hbar\Oph(q_{z}).\label{eq:P(h)}\end{equation}
where the symbol%
\footnote{For any $k\in\IR$, the symbol class $S^{k}(T^{*}X)$ denotes the
functions $a(x,\xi;\hbar)$ satisfying the following bounds: for any
multi-indices $\alpha,\beta\in\IN^{d}$,\[
\left|\partial_{x}^{\alpha}\partial_{\xi}^{\beta}a(x,\xi;\hbar)\right|\leq C_{\alpha,\beta}\,(1+|\xi|){}^{k-|\beta|}\quad\mbox{uniformly for }\hbar\in(0,1),\]
By $\Oph(q)$ we refer to a semiclassical quantization on $X$, with
the property that $\Oph(q)$ is symmetric if $q$ is real (see e.g
\cite[App. E]{EvZw}). The operators $\Oph(q)$ form the class $\Psi^{k}(X)$
of pseudodifferential operators (PDO). %
} $q_{z}\in S^{0}(T^{*}X)$ is now a function on the phase space $T^{*}X$,
and depends holomorphically on $z\in D(1/2,C\hbar)$, with $q_{z}=q_{1/2}+\cO(\hbar)$
\cite{Sjo00}. With a slight abuse, will call $q=q_{1/2}(x,\xi)$
{}``the damping function'' (a damping effectively occurs only at
phase space points where $q(x,\xi)<0$, while positive values of $q$
rather correspond to a {}``gain'' or {}``pumping''). 

In this framework, we are interested in the semiclassical distribution
of the generalized eigenvalues $z_{n}(\hbar)$ of the operator $P(\hbar,z)$
in the disk $D(1/2,C\hbar)$. Because we will specify this spectrum
at the scale $\hbar$, we may for convenience drop the term $\cO(\hbar^{2})$
in (\ref{eq:P(h)}), and rather study the spectrum of the $z$-independent
operator \begin{equation}
P(\hbar)=P_{0}(\hbar)+i\hbar\Oph(q),\qquad P_{0}(\hbar)\defeq-\frac{\hbar^{2}\Delta}{2}.\label{eq:P(h)-noz}\end{equation}
The principal symbol of this operator is $p_{0}(x,\xi)=\frac{|\xi|^{2}}{2}$,
it generates the geodesic flow $\Phi^{t}=\exp(tH_{p_{0}})$ on $T^{*}X$,
with unit speed on the energy shell $p_{0}^{-1}(1/2)=S^{*}X$.

\section{Semiclassical spectral distribution of $P(\hbar)$. General results}

The results we present in this section hold for arbitrary Riemannian
manifolds $(X,g)$. We skip the proofs.

\subsection{Weyl law for the real parts}

In the semiclassical limit, the number of eigenvalues of the selfadjoint
operator $P_{0}(\hbar)=-\frac{\hbar^{2}\Delta}{2}$ in the interval
$\left[\frac{1}{2}\pm\eps\right]\defeq\left[\frac{1}{2}-\eps,\frac{1}{2}+\eps\right]$
is approximately given by Weyl's law. When applying the perturbation
$i\hbar\Oph(q)=\cO(\hbar)$, it seems natural to expect that most
eigenvalues of $P_{0}(\hbar)$ in this interval will remain inside
a (complex) neighbourhood $\left[\frac{1}{2}\pm\eps\right]+\cO(\hbar)$.
One can indeed show that the {}``macroscopic'' spectral distribution
is given by the same Weyl's law as in the undamped case. 
\begin{thm}
\label{thm:Weyl law}\cite{MarMat84,Sjo00} For any $\eps>0$ small
enough (possibly depening on $\hbar$), the number of eigenvalues
of $P(\hbar)$ in the strip $\cS_{\eps}=\left\{ \frac{1}{2}-\eps\leq\Re z\leq\frac{1}{2}+\eps\right\} $
is semiclassically given by 

\[
\#\left\{ \Spec P(\hbar)\cap\cS_{\eps}\right\} =(2\pi\hbar)^{-d}\Vol\left\{ \rho\in T^{*}X,\,|p_{0}(\rho)-1/2|\leq\eps\right\} +\cO\left(\hbar^{-d+1}\right).\]
Here $\Vol$ corresponds to the symplectic volume in $T^{*}X$. This
estimate implies that, for $C>0$ large enough, \begin{equation}
\#\left\{ \Spec P(\hbar)\cap D(1/2,C\hbar)\right\} \asymp\hbar^{-d+1}.\label{eq:Weyl-hbar}\end{equation}

\end{thm}
This result is actually not so easy to prove. It uses some subtle
analytic Fredholm theory in order to relate the spectrum of $P(\hbar)$
with that of a simpler {}``comparison operator''.

\subsection{Restrictions for the imaginary parts (quantum decay rates)\label{sub:QDR}}

Eventhough the operator $P(\hbar)$ was inspired by the wave equation,
it is convenient for us to consider it instead as the generator of
a \emph{Schrödinger equation} \begin{equation}
i\hbar\partial_{t}v-P(\hbar)v=0.\label{eq:Schrod}\end{equation}
Any eigenvalue/vector $(z_{n},u_{n})$ of $P(\hbar)$ is thus associated
with a solution $v_{n}(x,t)=e^{-iz_{n}t/\hbar}u_{n}(x)$ of this equation,
and we will also call $\Im z_{n}/\hbar$ the \emph{quantum decay rate}%
\footnote{From (\ref{eq:tau-z}) we have $\Im z_{n}/\hbar=\Im\tau_{n}+\cO(\hbar)$,
so this denomination is (approximately) consistent with the one of
$\S$\ref{sub:intro}.%
} associated with $u_{n}$. 

First we remark that an eigenmode $u_{n}$ associated to an eigenvalue
$z_{n}\in D(1/2,C\hbar)$ is \emph{semiclassically microlocalized}
on $p_{0}^{-1}(1/2)=S^{*}X$. To state this property we need to introduce
energy cutoffs.
\begin{defn}
\label{def:cutoff}An energy cutoff will be function $\chi\in C_{c}^{\infty}\left(\left[1/2\pm\delta\right],[0,1]\right)$
for some $\delta>0$, and satisfies $\chi(s)=1$ near $s=1/2$. The
cutoff $\chi_{0}$ is said to be embedded in $\chi_{1}$, and we note
$\chi_{1}\succ\chi_{0}$, iff $\chi_{1}\equiv1$ near $\supp\chi_{0}$.
From the cutoff $\chi$ we construct the (pseudodifferential) cutoff
operator $\chi(P_{0}(\hbar))\in\Psi^{-\infty}(X)$.\end{defn}
\begin{prop}
\label{lem:Microloc} Take an energy cutoff $\chi$. Let $u=u(\hbar)$
be an eigenstate of $P(\hbar)$ with eigenvalue $z(\hbar)\in D(1/2,C\hbar)$.
Then, \[
\left\Vert \left(I-\chi(P_{0})\right)u\right\Vert =\cO(\hbar^{\infty})\left\Vert u\right\Vert .\]

\end{prop}
This localization property explains why only the neighbourhood of
$S^{*}X$ will be important for us.

\subsubsection{A factorization of the Schrödinger propagator}

An easy way to realize how the skew-adjoint subprincipal term $i\hbar\Oph(q)$
implies a {}``damping'' is to analyze the propagator for the Schrödinger
equation (\ref{eq:Schrod}). The following Proposition adapts a more
general result due to Rauch and Taylor \cite{RauTay75}. 
\begin{prop}
\label{pro:Vt=00003DUtBt}Assume $q\in C_{c}^{\infty}(T^{*}X)$, and
take $P(\hbar)$ as in (\ref{eq:P(h)-noz}). For any fixed $t\in\IR$,
decompose the Schrödinger propagator $V^{t}=e^{-itP/\hbar}$ into\begin{equation}
V^{t}=U^{t}\, B(t)\,,\;\mbox{where }U^{t}=e^{-itP_{0}/\hbar}\:\mbox{is the undamped propagator.}\label{eq:factorization}\end{equation}
The operator $B(t)$ is a PDO in $\Psi^{0}(X)$ of principal symbol
\[
b(t,\rho)=\exp\left\{ \int_{0}^{t}q\circ\Phi^{s}(\rho)\, ds\right\} =\exp\left\{ t\,\la q\ra_{t}(\rho)\right\} ,\]
 where the time averaged damping $\la q\ra_{t}$ generalizes (\ref{eq:a_T}).
\end{prop}
The factor $b(t,\rho)$ is the \emph{accumulated damping} along the
orbit $\left\{ \Phi^{s}(\rho),\:0\leq s\leq t\right\} $. If one starts
from a Gaussian wavepacket $u_{0}$ microlocalized at a point $\rho$,
then the state $V^{t}u_{0}$ is a wavepacket microlocalized at $\Phi^{t}(\rho)$,
and the above statement shows that the $L^{2}$ norms are related
through this factor: \[
\left\Vert u_{t}\right\Vert =\left\Vert u_{0}\right\Vert \left(b(t,\rho)+\cO(\hbar)\right).\]
Applying an energy cutoff allows to extend the factorization (\ref{eq:factorization})
damping functions $q\in S^{0}(T^{*}X)$ of noncompact support.
\begin{lem}
\label{lem:V^t chi_0}Consider $P(\hbar)$ with a damping function
$q\in S^{0}(T^{*}X)$. Take two embedded cutoffs $\chi_{1}\succ\chi_{0}$
supported in $\left[1/2\pm\delta\right]$, and the truncated damping
$\tilde{q}\defeq\chi_{1}(p_{0})\, q$. 

Then, for any $t\geq0$ fixed, one has \begin{equation}
V^{t}\,\chi_{0}(P_{0})=\tilde{V}^{t}\,\chi_{0}(P_{0})+\cO_{L^{2}\to L^{2}}(\hbar^{\infty})\,,\label{eq:factorization-chi0}\end{equation}
where $\tilde{V}^{t}$ is the propagator corresponding to $\tilde{P}=P_{0}+i\hbar\Oph(\tilde{q})$. 
\end{lem}
This identity uses the fact that the propagation $V^{t}$ does not
modify the energy localization properties.

\subsubsection{From propagator factorization to a resolvent bound}

We can now easily obtain a first constraint on the quantum decay rates.
Applying Proposition \ref{pro:Vt=00003DUtBt} to the function $\tilde{q}=\chi_{1}(p_{0})q$
of Lemma \ref{lem:V^t chi_0}, we get for any fixed $t\geq0$:\begin{align}
\left\Vert V^{t}\chi_{0}(P_{0})\right\Vert _{L^{2}\to L^{2}} & =\left\Vert \tilde{B}(t)\chi_{0}(P_{0})\right\Vert _{L^{2}\to L^{2}}+\cO(\hbar^{\infty})\nonumber \\
 & =\max_{T^{*}X}\tilde{b}(t)\chi_{0}(p_{0})+\cO_{t}(\hbar)\label{eq:bounds0}\\
 & \leq\exp\left\{ t\max_{\supp\chi_{0}\circ p_{0}}\la\tilde{q}\ra_{t}\right\} +\cO_{t}(\hbar).\nonumber \end{align}
On the second line we used the sharp Gårding (in)equality for the
operator $\tilde{B}(t)\in\Psi^{0}(X)$. The maximum $\max_{p_{0}^{-1}([1/2\pm\delta])}\la\tilde{q}\ra_{t}$$ $
decreases when $t\to\infty$ or when $\delta\to0$, and converges
to the asymptotic maximum on $S^{*}X$, \begin{equation}
q_{+}\defeq\lim_{t\to\infty}\max_{\rho\in S^{*}X}\la q\ra_{t}(\rho).\label{eq:q+-}\end{equation}
One similarly defines an asymptotic minimum $q_{-}$. We then get
the following norm bounds for the propagator.
\begin{prop}
Fix $\eps>0$. If the energy cutoff $\chi_{0}$ has a small enough
support, and $T_{\eps}>0$ is large enough, then for any (fixed) $t\geq T_{\eps}$
and any small enough $\hbar>0$, the following bounds holds: \begin{equation}
\left\Vert V^{t}\chi_{0}(P_{0})\right\Vert _{L^{2}\to L^{2}}\leq e^{\left(q_{+}+\eps\right)t}\,,\qquad\left\Vert V^{-t}\chi_{0}(P_{0})\right\Vert _{L^{2}\to L^{2}}\leq e^{\left(-q_{-}+\eps\right)t}\,.\label{eq:V^t chi-bound}\end{equation}

\end{prop}
From there one easily obtains the following resolvent and spectral
bounds.
\begin{thm}
\label{thm:extremal eigenv}\cite{Leb93,Sjo00}Take any $\eps>0$.
Then, there exists $\hbar_{\eps},C_{\eps}>0$ such that for $\hbar<\hbar_{\eps}$,
the following resolvent estimate holds:\begin{equation}
\forall z\in D(1/2,C\hbar)\setminus\left\{ z,\:\Im z/\hbar\in[q_{-}-\eps,q_{+}+\eps]\right\} ,\qquad\left\Vert \left(P(\hbar)-z\right)^{-1}\right\Vert \leq\frac{C_{\eps}}{\hbar}.\label{eq:resolvent-general}\end{equation}
As a consequence, for $\hbar<\hbar_{\eps}$ all eigenvalues $z_{n}\in D(1/2,C\hbar)$
of $P(\hbar)$ satisfy \[
\frac{\Im z_{n}}{\hbar}\in\left[q_{-}-\eps,q_{+}+\eps\right].\]

\end{thm}
In the case of the damped wave equation ($q(x,\xi)=-a(x)$) and in
a situation of geometric control ($a_{-}=-q_{+}>0$), the resolvent
estimate (\ref{eq:resolvent-general}) can be used to show the uniform
exponential energy decay (\ref{eq:decay-GCC}) \cite{Hit03}.

\subsection{Questions on the spectral distribution\label{sub:Questions}}

So far we showed that the quantum decay rates are bounded by the asymptotic
extrema of the time-averaged damping. The following questions were
raised in \cite{Sjo00,AschLeb,Anan10} concerning their semiclassical
distribution.
\begin{enumerate}
\item What are the possible accumulation points of the quantum decay rates
when $\hbar\to0$? In particular, are there sequences of decay rates
$(\Im z(\hbar)/\hbar)_{\hbar\to0}$ converging to the extremal values
$q_{\pm}$?
\item Do the quantum decay rates admit an asymptotic distribution when $\hbar\to0$?
Namely, for a given interval $I\subset[q_{-},q_{+}]$ and $1\gg\eps(\hbar)\gg\hbar$,
does the ratio \[
\frac{\#\left\{ z\in\Spec P(\hbar),\ |\Re z-1/2|\leq\eps,\:\frac{\Im z}{\hbar}\in I\right\} }{\#\left\{ z\in\Spec P(\hbar),\ |\Re z-1/2|\leq\eps\right\} }\]
have a limit when $\hbar\to0$? Is this limit distribution related
with the value distributions of the averages $\la q\ra_{t}$?
\end{enumerate}

\section{Spectral estimates on Anosov manifolds}

The above questions are open in general. In order to get more precise
informations on the spectrum of $P(\hbar)$, one needs to make specific
assumptions on the geodesic flow on $X$. For instance, the case of
a completely integrable dynamics has been considered by Hitrik-Sjöstrand
in a sequence of papers (see e.g. \cite{HitSjo08}and reference therein).
The case of nearly-integrable dynamics including KAM invariant tori
has been studied by Hitrik-Sjöstrand-V\~u Ng\d{o}c \cite{HSVN07}.
In these cases, one can transform the Hamiltonian flow into a \emph{normal
form} near each invariant torus, which leads to a precise description
of the spectrum {}``generated'' by this torus. A Weyl law for the
quantum decay rates was recently obtained in \cite{HitSjo11} (for
skew-adjoint perturbations $i\theta(\hbar)\Oph(q)$, with $\theta(\hbar)\ll\hbar$).
On the other hand, Asch-Lebeau \cite{AschLeb} addressed Question
1 for the case of the 2-dimensional standard sphere. They show that,
if the damping function $q$ has real analytic real and imaginary
parts, there is generically a \emph{spectral gap} $\gamma>0$: for
$\hbar$ small enough, all eigenvalues $z_{n}\in D(1/2,C\hbar)$ have
quantum decay rates \[
\Im z_{n}/\hbar\in\left[q_{-}+\gamma,q_{+}-\gamma\right].\]
The proof proceeds by applying a complex canonical transformation,
such that the real part of the pulled-back damping function takes
values in the above interval. In this case, the range of the quantum
decay rates is strictly smaller than the range of classical decay
rates. 

We will see below that such a spectral gap may also occur in the case
of Anosov geodesic flows, thanks to a different mechanism, namely
a \emph{hyperbolic dispersion} property due to the instability of
the classical flow (see Thms \ref{thm:pressure-resolvent} and \ref{thm:thickness-bound}
below). 

In the next section we recall the definition and properties of Anosov
manifolds.

\subsection{A short reminder on Anosov manifolds \cite{KatHas95}\label{sub:Anosov-reminder}}

\begin{figure}
\includegraphics[scale=0.5]{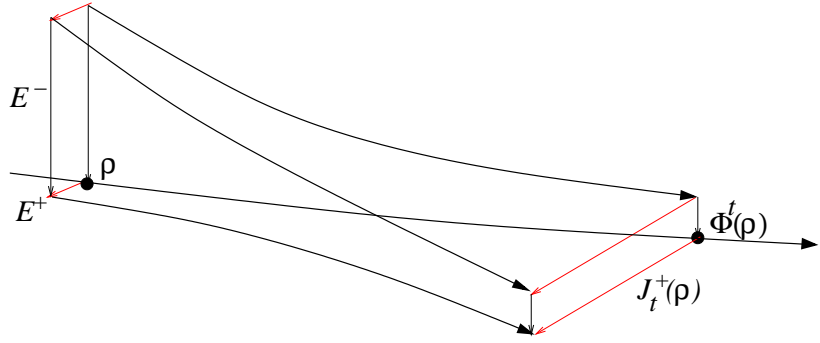}\caption{Structure of the Anosov flow near an orbit $\Phi^{t}(\rho)$}
\end{figure}

At the {}``antipode'' of the completely integrable case, one finds
the {}``strongly chaotic'' flows, namely the Anosov (or uniformly
hyperbolic) flows. Uniform hyperbolicity means that at each point
$\rho\in S^{*}X$ there exists a splitting of the tangent space, \[
T_{\rho}S^{*}X=\IR H_{p}(\rho)\oplus E^{+}(\rho)\oplus E^{-}(\rho),\]
where $H_{p}(\rho)$ is the Hamiltonian vector field, $E^{\pm}(\rho)$
are the unstable and stable subspaces at the point $\rho$; they both
have dimension $d-1$, and are uniformly transverse to e.o.. The families
$\left\{ E^{\pm}(\rho),\,\rho\in S^{*}X\right\} $ form the unstable/stable
distributions, they are invariant w.r.to the flow, Hölder continuous,
and are characterized by the following property: there exists $C,\lambda>0$
such that\begin{equation}
\forall\rho\in S^{*}X,\:\forall v\in E^{\mp}(\rho),\:\forall t>0,\quad\left\Vert d\Phi_{\rho}^{\pm t}v\right\Vert \leq C\, e^{-\lambda t}\,\left\Vert v\right\Vert .\label{eq:hyperb}\end{equation}
As was shown by Hadamard, such a geodesic flow is obtained if the
manifold $(X,g)$ has a negative sectional curvature. The long time
properties of such a strongly chaotic flow are well understood. In
particular, Hopf proved that this geodesic flow is \emph{ergodic}
w.r.to the Liouville measure on $S^{*}X$. 
\begin{defn}
Call $\mu_{L}$ the normalized Liouville measure on $S^{*}X$. Then,
the geodesic flow $\Phi^{t}$ on $S^{*}X$ is ergodic (w.r.to $\mu_{L}$)
iff, for any continuous function $q$ on $S^{*}X$, \[
\mbox{for }\:\mu_{L}-\mbox{almost every }\rho\in S^{*}X,\qquad\lim_{t\to\infty}\la q\ra_{t}(\rho)=\bar{q}\defeq\mu_{L}(q).\]

\end{defn}
The following quantities provide quantitative measures of the hyperbolicity:\begin{align}
\mbox{the maximal expansion rate}\quad & \lambda_{\max}\defeq\lim_{t\to\infty}\sup_{\rho\in S^{*}X}\frac{1}{t}\log\left\Vert d\Phi_{\rho}^{t}\right\Vert ,\label{eq:lambda_max}\\
\mbox{the unstable Jacobian}\quad & J^{+}(\rho,t)=\det\left(d\Phi^{t}\rest_{E^{+}(\rho)}\right),\quad t>0,\nonumber \\
\mbox{and its infinitesimal version }\quad & \varphi^{+}(\rho)\defeq\frac{d}{dt}\log J^{+}(\rho,t)\restriction_{t=0},\label{eq:infinit-jacob}\\
\mbox{the minimal expansion rate}\quad & \nu_{\min}\defeq\frac{1}{d-1}\lim_{t\to\infty}\min_{\rho\in S^{*}X}\frac{1}{t}\log J^{+}(\rho,t)\,.\label{eq:nu_min}\end{align}
The Jacobian depends on the choice of norms on the spaces $T_{\rho}S^{*}X$.
However, its long time asymptotics is independent of this choice.
In variable curvature these rates are related by $\lambda_{\max}\geq\nu_{\min}\geq\lambda>0$.
In particular, if the positive Lyapunov exponents are not all equal,
one has $\nu_{\min}<\lambda_{\max}$. 

A particular class of Anosov manifolds consists in quotients of the
$d$-dimensional hyperbolic space $\IH^{d}$ by co-compact subgroups
of its isometry group. These manifolds have a constant curvature%
\footnote{Usually one normalizes the metrics on $\IH^{d}$ so that the curvature
is $-1$. We prefer to keep track of the curvature in our notations.%
} $-\Lambda^{2}$. Using the natural norms on $T_{\rho}S^{*}X$, one
then has \[
J^{+}(\rho,t)=e^{t(d-1)\Lambda},\quad\varphi^{+}(\rho)=\Lambda(d-1),\quad\lambda_{\max}=\nu_{\min}=\Lambda.\]
In this case the contraction/expansion are both homogeneous (independent
of the point $\rho$) and isotropic (independent in the direction).

Let us finally notice that the study of the operator $P(\hbar)$ on
an Anosov manifold belongs to the field of {}``quantum chaos''.
The methods we will use below occur in various problems of this field,
e.g. the study of eigenstates of the Laplacian $P_{0}(\hbar)$ on
such manifolds \cite{Zel09}.

\subsection{Fractal Weyl upper bounds for the quantum decay rates}

In this section we address Question 2, that is the asymptotic distribution
of the quantum decay rates, for the case of Anosov manifolds.

\subsubsection{Typical quantum decay rates for ergodic flows }

A basic property of the geodesic flow on any Anosov manifold is ergodicity.
Sjöstrand has studied the asymptotic disribution of the quantum decay
rates for any ergodic geodesic flow \cite{Sjo00}, and obtained the
following result. 
\begin{thm}
\label{thm:essential support}\cite{Sjo00} Assume the geodesic flow
on $S^{*}X$ is ergodic w.r.to the Liouville measure. Then, For any
$C>0,\,\eps>0$, one has as $\hbar\to0$, \[
\#\left\{ z\in\Spec P(\hbar)\cap D(1/2,C\hbar),\:\frac{\Im z}{\hbar}\not\in[\bar{q}-\eps,\bar{q}+\eps]\right\} =o(\hbar^{-d+1}).\]

\end{thm}
Comparing this bound with the Weyl law (\ref{eq:Weyl-hbar}) shows
that, in the semiclassical limit, \emph{almost all} the quantum decay
rates are close to the phase space average $\bar{q}$. The latter
could be called the \emph{typical value} for quantum decay rates,
while values at finite distance from $\bar{q}$ are \emph{atypical}.
The asymptotic distribution is simply the delta measure $\delta_{\bar{q}}$.
\begin{rem}
\label{rem:QE-link}As was noticed in \cite{Anan10}, this concentration
result could be seen as a {}``nonperturbative version'' of \emph{quantum
ergodicity}. The latter property states that \emph{almost all} the
eigenstates $u_{n}^{0}$ of $P_{0}(\hbar)$ with eigenvalues $z_{n}^{0}\in D(1/2,C\hbar)$
satisfy $\la u_{n}^{0},\Oph(q)u_{n}^{0}\ra=\bar{q}+o_{\hbar\to0}(1)$.
A naive perturbation theory argument would predict that, when switching
on $i\hbar\Oph(q)$, the eigenvalues $z_{n}^{0}$ move to $z_{n}=z_{n}^{0}+i\hbar\la u_{n}^{0},\Oph(q)u_{n}^{0}\ra+\cO(\hbar^{2})$,
so that the quantum decay rates $\Im z_{n}/\hbar=\la u_{n}^{0},\Oph(q)u_{n}^{0}\ra+\cO(\hbar)$,
the RHS being equal to $\bar{q}+o(1)$ for almost all $n$. Of course,
this argument does not apply because the perturbation $i\hbar\Oph(q)$
is much stronger than the mean spacing between successive eigenvalues.
\end{rem}
To prove the above theorem one relates the counting of quantum decay
rates with the value distribution of the quantum averages $\la q\ra_{t}$
on $S^{*}X$, that is the volumes\begin{equation}
\mu_{L}\left\{ \rho\in S^{*}X,\;\la q\ra_{t}(\rho)\geq\alpha\right\} ,\quad\alpha\in\IR,\; t>0.\label{eq:volume}\end{equation}
Indeed, the main intermediate result in the proof is the bound \begin{equation}
\#\left\{ z\in\Spec P(\hbar)\cap D(1/2,C\hbar),\:\frac{\Im z}{\hbar}\geq\alpha\right\} \leq C_{t,\delta}\:\hbar^{-d+1}\,\mu_{L}\left\{ \rho\in S^{*}X,\;\la q\ra_{t}(\rho)\geq\alpha-\delta\right\} ,\label{eq:bound1}\end{equation}
which holds for any fixed $t>0$ and $\delta>0$ (this bound holds
independently of the ergodicity assumption). Ergodicity then implies
that the value distribution of $\la q\ra_{t}$ converges to $\delta_{\bar{q}}$
when $t\to\infty$; in particular, for any $\alpha>\bar{q}$ the volume
(\ref{eq:volume}) decays to zero when $t\to\infty$. We have then
obtained the bound $o(\hbar^{-d+1})$ for the quantum decay rates
$\geq\alpha>\bar{q}$. The case of values $\leq\alpha<\bar{q}$ is
treated analogously. $\hfill\square$

\subsubsection{Large deviation estimates for Anosov flows}

For an Anosov flow, one has more precise estimates for the volumes
(\ref{eq:volume}), which can then induce sharper bounds on the number
of the atypical quantum decay rates. These volume estimate take the
form of \emph{large deviation} estimates. Let us introduce some notations.
We call $\cM$ the set of $\Phi^{t}$-invariant probability measures
on $S^{*}X$. For each such measure $\mu$, we denote by $h_{KS}(\mu)$
its Kolmogorov-Sinai entropy: this is a nonnegative number, which
measures the complexity of a $\mu$-typical trajectory \cite{KatHas95}.
Then, we define the rate function $\tilde{H}:\IR\to\IR$ as follows:\begin{equation}
\forall s\in\IR,\qquad\tilde{H}(s)\defeq\sup\left\{ h_{KS}(\mu)-\mu(\varphi^{+}),\;\mu\in\cM,\:\mu(q)=s\right\} ,\label{eq:tildeH(s)}\end{equation}
where we recall that $\varphi^{+}$ is the infinitesimal unstable
Jacobian (\ref{eq:infinit-jacob}). We are now ready to state our
large deviation result.
\begin{thm}
\cite{Kif90}Assume the geodesic flow on $S^{*}X$ is Anosov. Then,
for any closed interval $I\subset\IR$ and for any $q\in C^{\infty}(S^{*}X)$,
the time averages $ $$\la q\ra_{t}$ satisfy \begin{equation}
\limsup_{t\to\infty}\frac{1}{t}\log\mu_{L}\left\{ \rho\in S^{*}X,\,\la q\ra_{t}(\rho)\in I\right\} \leq\sup_{s\in I}\tilde{H}(s).\label{eq:large-deviation-variable}\end{equation}

\end{thm}
The rate function $\tilde{H}$ is continuous on $[q_{-},q_{+}]$,
smooth and strictly concave on $(q_{-},q_{+})$, negative except at
the point $\tilde{H}(\bar{q})=0$, satisfies $\tilde{H}(s)\geq-\sup_{\mu\in\cM}\mu(\varphi^{+})$
for $s\in[q_{-},q_{+}]$, and is equal to $-\infty$ outside $[q_{-},q_{+}]$.
As a consequence of (\ref{eq:large-deviation-variable}), for any
$\alpha\geq\bar{q}$ and $\eps>0$ arbitrary small, there exists $T_{\alpha,\eps}>0$
such that \begin{equation}
\forall t\geq T_{\alpha,\eps},\quad\mu_{L}\left\{ \rho\in S^{*}X,\,\la q\ra_{t}(\rho)\geq\alpha\right\} \leq e^{t\left(\tilde{H}(\alpha)+\eps\right)}.\label{eq:large-deviation2}\end{equation}
Due to the negativity of $\tilde{H}(\alpha)$ for $\alpha\neq\bar{q}$,
we see that the probability of $\la q\ra_{t}$ taking atypical values
decays exponentially when $t\to\infty$.

\subsubsection{Fractal Weyl upper bounds on Anosov manifolds\label{sub:Fractal-Weyl-upper}}

Using these large deviation estimates, Anantharaman \cite{Anan10}
improved Thm \ref{thm:essential support} by letting the averaging
time grow with $\hbar$ in a controlled way. The optimal time is the
\emph{Ehrenfest time} \begin{equation}
T=T_{Ehr}=\left(1-2\eps\right)\frac{\log1/\hbar}{\lambda_{\max}},\label{eq:Ehrenfest}\end{equation}
where $\lambda_{\max}$ is the largest expansion rate (\ref{eq:lambda_max})
and $\eps>0$ arbitrary small. What is the signification of this time?
For any $f\in S^{-\infty}(T^{*}X)$ supported in an $\eps$-neighbourhood
of $S^{*}X$, the classically evolved observable $f\circ\Phi^{t}$
remains in the {}``good'' symbol class%
\footnote{For any $k\in\IR$, $\delta\in[0,1/2)$, the symbol class $S_{\delta}^{k}(T^{*}X)$
consists of functions $g(x,\xi;\hbar)$ which may become more and
more singular when $\hbar\to0$, but in a controlled way:\[
\forall\alpha,\beta\in\IN^{d},\forall\rho\in T^{*}X,\quad\left|\partial_{x}^{\alpha}\partial_{\xi}^{\beta}g(\rho)\right|\leq C_{\alpha}\,\hbar^{-\delta(|\alpha|+|\beta|)}\,(1+|\xi|)^{k-|\beta|}.\]
In this class one can still use pseudodifferential calculus, and the
expansions in powers of $\hbar$ make sense \cite[Sec. 4.3]{EvZw}. %
} $S_{1/2-\eps}^{-\infty}(T^{*}X)$ uniformly for times $|t|\leq T_{Ehr}/2$.
In turn, the \emph{symmetric} averages\begin{equation}
\la f\ra_{t,sym}=\frac{1}{t}\int_{-t/2}^{t/2}f\circ\Phi^{s}\, ds\label{eq:symm-average}\end{equation}
belong to $S_{1/2-\eps}^{-\infty}(T^{*}X)$ for $|t|\leq T_{Ehr}$,
and this time is sharp. 

Let us insist on the fact that controlling the time evolution up to
times $t\asymp\log1/\hbar$ is a crucial ingredient in order to obtain
refined spectral estimates on Anosov manifolds. This will also be
the case in Sections \ref{sub:pressure-criterium} and \ref{sub:thickness-condition}
when proving hyperbolic dispersion estimates and spectral gaps.

The pseudodifferential calculus on $\Psi_{1/2-\eps}^{2}(X)$ allows
to extend the validity of the bound (\ref{eq:bound1}) up to the Ehrenfest
time. The large deviation estimate (\ref{eq:large-deviation2}) then
leads to the following \emph{fractal Weyl upper bound} for the number
of atypical quantum decay rates.
\begin{thm}
\label{thm:fractal-Weyl-flow}\cite{Anan10}Assume the geodesic flow
on $S^{*}X$ is Anosov. Then, for any $\alpha\geq\bar{q}$ and $\eps>0$,
one has for $\hbar$ small enough \begin{equation}
\#\left\{ z\in\Spec P(\hbar)\cap D(1/2,C\hbar),\:\frac{\Im z}{\hbar}\geq\alpha\right\} \leq\hbar^{-\frac{\tilde{H}(\alpha)}{\lambda_{\max}}+(1-d)-\eps},\label{eq:fractal-upper-variable}\end{equation}
where $\tilde{H}$ is the rate function (\ref{eq:tildeH(s)}). A similar
expression holds when counting quantum decay rates smaller than $\alpha'\leq\bar{q}$.
\end{thm}
Since $-\tilde{H}(\alpha)>0$ for all $\alpha>\bar{q}$, this upper
bound improves the bound $o(\hbar^{-d+1})$ of Thm. \ref{thm:essential support}
by a fractional power of $\hbar$ (the name {}``fractal Weyl upper
bound'' takes its origin in the counting of resonances of chaotic
scattering systems \cite{SjoZwo07}). For any $\alpha\in\left[\bar{q},q_{+}\right]$
one has $\tilde{H}(\alpha)\geq-\sup_{\mu}\mu(\varphi^{+})\geq-\lambda_{\max}(d-1)$,
so the exponent of $\hbar$ in (\ref{eq:fractal-upper-variable})
is negative, allowing the presence of (many) quantum decay rates arbitrary
close to $q_{+}$. 

These fractal upper bounds are not expected to be sharp in general
\cite{Anan10}. In the next sections we will be able, under certain
conditions, to exclude the possibility of quantum decay rates near
$q_{+}$ (or near $q_{-}$).

\subsection{A pressure criterium for a spectral gap\label{sub:pressure-criterium}}

The result of Thm \ref{thm:extremal eigenv}, namely the fact that
all quantum decay rates belong to the interval $\left[q_{-}-o(1),q_{+}+o(1)\right]$,
was obtained by studying the norm of the propagator $V^{t}=e^{-itP/\hbar}$
for times $|t|\gg1$ independent of $\hbar$. Indeed, the sequence
of equalities (\ref{eq:bounds0}) directly leads to the upper bound
\begin{equation}
\Im z_{n}/\hbar\leq q_{+}+\eps.\label{eq:bound-sup}\end{equation}
The crucial semiclassical ingredient in (\ref{eq:bounds0}) is the
second equality connecting the $L^{2}$ norm of the PDO $\tilde{B}(t)\chi(P_{0})$
with the supremum of its (principal) symbol. Hence, one cannot hope
to improve the bound (\ref{eq:bound-sup}) as long as $\tilde{B}(t)\chi(P_{0})$
remains a {}``good'' PDO, which is the case if the time $t\leq T_{Ehr}/2$.
To improve on (\ref{eq:bound-sup}) we will investigate the propagator
$V^{t}$ for {}``large logarithmic times'', namely \begin{equation}
t\sim\cK\log1/\hbar,\quad\mbox{with }\cK>0\;\mbox{ large (but independent of \mbox{\ensuremath{\hbar})}.}\label{eq:large-logarithmic}\end{equation}
For such times, the function $\tilde{b}(t)\chi(p_{0})$ oscillates
on scales much smaller than $\hbar$, so it cannot belong to any decent
symbol class. As a result, the norm of $\tilde{B}(t)\chi(P_{0})$
is a priori unrelated with that function. However, using the hyperbolicity
of the flow one can prove an upper bound for the propagator $V^{t}$
in terms of a certain \emph{topological pressure} depending on the
damping and the hyperbolicity, and then obtain a constraint on the
quantum decay rates.

Before stating our bound, let us first introduce the notion of topological
pressure associated with the flow $\Phi^{t}$ on $S^{*}X$. For any
observable $f\in C(S^{*}X)$, the pressure $\cP(f)=\cP(f,\Phi^{t}\rest_{S^{*}X})$
can be defined by\begin{equation}
\cP(f)\defeq\sup\left\{ h_{KS}(\mu)+\mu(f),\;\mu\in\cM\right\} .\label{eq:pressure}\end{equation}
Notice that the definition (\ref{eq:tildeH(s)}) of the rate function
$\tilde{H}(s)$ is very similar. The pressure is also given by the
growth rate of weighted sums over long closed geodesics%
\footnote{The sum over long closed geodesics is analogous to a \emph{partition
function} in statistical mechanics, with the time $T$ corresponding
to the volume of the system. This explains why the growth rate is
called a {}``pressure''.%
}: \[
\cP(f)=\lim_{T\to\infty}\frac{1}{T}\log\sum_{T\leq|\gamma|\leq T+1}e^{\int_{\gamma}f}.\]
The pressure appearing in the next theorem is associated with the
function $f=q-\varphi^{+}/2$, where $\varphi^{+}$ is the infinitesimal
unstable Jacobian (\ref{eq:infinit-jacob}), it thus mixes the damping
and hyperbolicity.
\begin{thm}
\cite{Sche10}\label{thm:pressure-resolvent}Let $X$ be an Anosov
manifold, and $q\in S^{0}(T^{*}X)$ a damping function. 

Assume the following (purely classical) inequality holds: \begin{equation}
\cP(q-\varphi^{+}/2)<q_{+}.\label{eq:good-bound}\end{equation}
Then, for any $\eps>0$, there exists $N>0$ such that, for $\hbar$
small enough, the following resolvent bounds:\begin{equation}
\forall z\in D(1/2,C\hbar)\cap\left\{ \Im z/\hbar\geq\cP(q-\varphi^{+}/2)+\eps\right\} ,\qquad\left\Vert \left(P(\hbar)-z\right)^{-1}\right\Vert \leq\hbar^{-N}.\label{eq:resolvent-pressure}\end{equation}
As a consequence, all quantum decay rates for $z_{n}(\hbar)\in D(1/2,C\hbar)$
satisfy \begin{equation}
\frac{\Im z_{n}(\hbar)}{\hbar}\leq\cP(q-\varphi^{+}/2)+\eps.\label{eq:gap}\end{equation}
in particular we have a spectral gap.
\end{thm}
\begin{figure}
\includegraphics[scale=0.5]{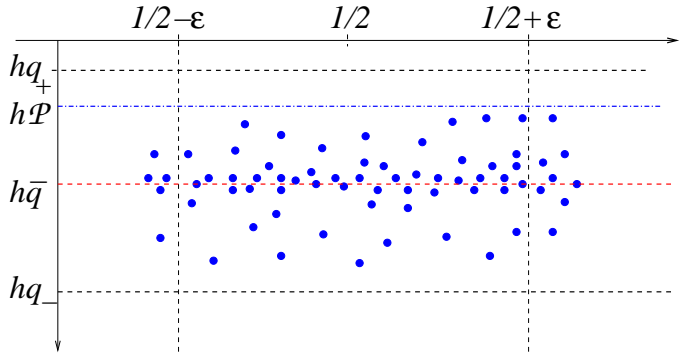}\caption{Spectral gap in case the pressure $\cP(q-\varphi^{+}/2)<q_{+}$.}
\end{figure}
In the next subsection we present situations for which the inequality
(\ref{eq:good-bound}) is satisfied. The proof of this theorem was
inspired by a similar result in the case of resonances in chaotic
scattering \cite{NZ2,NZ3}. We give some hints of the proof in $\S$\ref{sub:splitting-paths}. 

Let us now apply this bound to the damped wave equation, that is take
$q(x,\xi)=-a(x)$. 
\begin{cor}
\label{cor:pressure-DWE}\cite{Sche10} Consider the DWE on $X$ with
nontrivial damping $a\geq0$. Assume that the GCC fails ($a_{-}=0$),
but that the condition $\cP(-a-\varphi^{+}/2)<0$ holds true. 

Then the DWE has a spectral gap, and the energy decays exponentially
for regular initial data, as in (\ref{eq:decay-loss}).
\end{cor}

\subsubsection{Conditions for a spectral gap\label{sub:Schenck-stab}}

The inequality (\ref{eq:good-bound}) can hold only if the damping
function $q$ varies {}``sufficiently'' around its average value
$\bar{q}$. Indeed, one always has $\cP(-\varphi^{+}/2)>0$ for an
Anosov flow, so by adding a constant dampint $q\equiv\bar{q}$ one
gets \[
\cP(\bar{q}-\varphi^{+}/2)=\bar{q}+\cP(-\varphi^{+}/2)>\bar{q}=q_{+}.\]
The pressure depends continuously of $q$ (in the $C^{0}$ topology),
so if $q=\bar{q}+\delta q$ with $\left\Vert \delta q\right\Vert _{C^{0}}$
small, the inequality (\ref{eq:good-bound}) will not be satisfied. 

Let us consider the case of the DWE as in Corollary \ref{cor:pressure-DWE}.
The failure of the GCC is equivalent to the fact that the set of undamped
trajectories $K$ (\ref{eq:undamped-set}) is nonempty. This set is
flow-invariant and closed, so one can define the pressure $\cP(-\varphi^{+}/2,\Phi^{t}\rest_{K})$
associated with the flow $\Phi^{t}$ restricted on $K$. It was shown
in \cite{Sche11} that if we multiply the damping $a(x)$ by a large
constant $C>0$ (this does not modify the set $K$), then \[
\lim_{C\to\infty}\cP(-Ca-\varphi^{+}/2)=\cP(-\varphi^{+}/2,\Phi^{t}\rest_{K}).\]
If the set $K$ is \emph{thin enough,} the pressure $\cP(-\varphi^{+}/2,\Phi^{t}\rest_{K})$
is negative (for instance, if $K$ consists in a single closed geodesic,
the pressure is equal to the average of $-\varphi^{+}/2$ along the
geodesic, which is negative). In that case, the above limit shows
that the pressure $\cP(-Ca-\varphi^{+}/2)$ is also negative if $C\gg1$.

\subsubsection{Proof of the pressure bound: decomposing into {}``symbolic paths''\label{sub:splitting-paths}}

In this section we explain the strategy of proof of Thm \ref{thm:pressure-resolvent}.
As mentioned before, we want to bound the norm of the propagator for
large logarithmic times. We will prove the following 
\begin{lem}
Choose $\eps>0$. Then, for an energy cutoff $\chi_{0}$ of small
enough support, and a large enough constant $\cK>0$, for $\hbar<\hbar_{\eps}$
one has the bound \begin{equation}
\left\Vert V^{t}\chi_{0}(P_{0})\right\Vert \leq e^{t\left(\cP(q-\varphi^{+}/2)+\eps\right)},\quad t\sim\cK\log1/\hbar.\label{eq:pressure-bound}\end{equation}

\end{lem}
From there, obtaining the resolvent estimate (\ref{eq:resolvent-pressure})
is relatively straightforward.
\begin{proof}
Our aim is to bound the norm $\left\Vert V^{t}u_{0}\right\Vert $
for a \emph{normalized} state $u_{0}$ microlocalized in $p_{0}^{-1}\left(\left[1/2\pm\delta\right]\right)$,
and a time $t\sim\cK\log1/\hbar$. We may assume that $u_{0}$ is
microlocalized in an open subset $W$ of diameter $\leq\delta$. We
will first decompose $u_{0}$ in an adapted basis of local {}``momentum''
modes $\left\{ e_{\eta},\:\eta\in\IR^{d},\:|\eta|\leq\delta\right\} $.
Here the parameter $\eta$ is the {}``momentum coordinate'' in a
Darboux coordinate frame $\left\{ (y,\eta),\,|y|\leq\delta,\,|\eta|\leq\delta\right\} $
in $W$, chosen such that $\eta_{1}=p_{0}-1/2$ represents the energy,
$y_{1}$ the time, and the {}``momentum'' Lagrangian leaves $\Lambda_{\eta_{0}}=\left\{ (y,\eta_{0})\right\} $
are close the unstable foliation. The momentum states $e_{\eta}$
are Lagrangian states supported by the leaves $\Lambda_{\eta}$, of
norms $\cO(1)$. The decomposition of $u_{0}$ in this {}``basis''
reads \begin{equation}
u_{0}=\int_{|\eta|\leq\delta}\hat{u}_{0}(\eta)\, e_{\eta}\,\frac{d\eta}{(2\pi\hbar)^{d/2}}+\cO(\hbar^{\infty}),\label{eq:Fourier-expansion}\end{equation}
where the factor $(2\pi\hbar)^{-d/2}$ ensures that $\left\Vert \hat{u}_{0}\right\Vert _{L^{1}}=\cO(1).$
The reason for this decomposition is that we are able to precisely
control the evolution of each individual state $e_{\eta}$, up to
times $t\sim\cK\log1/\hbar$. The state $V^{t}e_{\eta}$ is a Lagrangian
state supported on the Lagrangian leaf $\Phi^{t}(\Lambda_{\eta})$.
Due to the hyperbolicity of the flow, when $t$ grows the leaf $\Phi^{t}(\Lambda_{\eta})$
expands exponentially fast along the unstable directions, and converges
to the unstable foliation. To analyze the state $V^{t}e_{\eta}$,
we can view as the combination of many local Lagrangian states associated
with bounded pieces of $\Phi^{t}(\Lambda_{\eta})$. A convenient bookkeeping
consists in using a partition of unity in $p_{0}^{-1}\left(\left[1/2\pm\delta\right]\right)$.
One first covers this region by a finite family of open subsets $\left\{ W_{j},\, j=1,\ldots,J\right\} $
of diameters $\leq\delta$, and then constructs a smooth partition
of unity $\left\{ \pi_{j}\in C_{c}^{\infty}(W_{j},[0,1]),\, j=1,\ldots,J\right\} $
adapted to this cover: $\sum_{j=1}^{J}\pi_{j}=1$ near $p_{0}^{-1}\left([1/2\pm\delta]\right)$.
This partition is quantized into $\left\{ \Pi_{j}\defeq\Oph(\pi_{j}),\, j=1,\ldots,J\right\} $,
such that, for any energy cutoff $\chi$ any supported in $[1/2\pm\delta]$,
\begin{equation}
\sum_{j=1}^{J}\Pi_{j}\,\chi(P_{0})=\chi(P_{0})+\cO_{L^{2}\to L^{2}}(\hbar^{\infty}).\label{eq:quantum_partition}\end{equation}
We use this quantum partition to split the propagator $V^{n}$ into
{}``symbolic paths'': \begin{equation}
V^{n}e_{\eta}=\sum_{\alpha_{1},\ldots,\alpha_{n}=1}^{J}V_{\alpha_{1}\cdots\alpha_{n}}e_{\eta}+\cO(\hbar^{\infty}),\quad V_{\alpha_{1}\cdots\alpha_{n}}=\Pi_{\alpha_{n}}V\Pi_{\alpha_{n-1}}\cdots V\Pi_{\alpha_{1}}V.\label{eq:decompo}\end{equation}
For each symbolic sequence $\balpha=\alpha_{1}\cdots\alpha_{n}$,
the state $V_{\balpha}e_{\eta}$ is a Lagrangian state supported on
the Lagrangian leaf $\Lambda_{\eta}^{\balpha}$ obtained from $\Lambda_{\eta}$
by a sequence of $n$ evolutions by $\Phi^{1}$ and truncations on
$W_{\alpha_{k}}$. The full evolved leaf is given by $\Phi^{t}(\Lambda_{\eta})=\bigcup_{|\balpha|=n}\Lambda_{\eta}^{\balpha}$.
If $\Lambda_{\eta}^{\balpha}$ is empty $V_{\balpha}e_{\eta}=\cO(\hbar^{\infty})$.
Otherwise ($\balpha$ {}``admissible''), $\Lambda_{\eta}^{\balpha}$
is a leaf of diameter $\leq\delta$, close to the unstable foliation.
One can compute the amplitude of the Lagrangian state $V_{\balpha}e_{\eta}$
to any order in $\hbar$. It is governed by both the accumulated damping\emph{
}and accumulated instability along the {}``path'' $\balpha$, leading
to the norm estimate \begin{equation}
\left\Vert V_{\balpha}e_{\eta}\right\Vert \leq C\, b(\balpha)\, J^{+}(\balpha)^{-1/2},\qquad\mbox{where }\label{eq:hyper-1}\end{equation}
\begin{equation}
b(\balpha)=\prod_{k=1}^{n}b(\alpha_{k}),\qquad b(j)=\max_{\rho\in W_{j}}b(1,\rho),\qquad\mbox{and similarly for }\:\left(J^{+}\right)^{-1/2}.\label{eq:b(alpha)}\end{equation}
Applying the triangular inequality to (\ref{eq:decompo}), we obtain
a bound on the norm of $V^{n}e_{\eta}$ which can be related with
the topological pressure: \begin{equation}
\left\Vert V^{n}e_{\eta}\right\Vert \leq C\sum_{\balpha\, admis.}^{J}b(\balpha)\, J^{+}(\balpha)^{-1/2}\leq\exp\left\{ n\left(\cP(q-\varphi^{+}/2)+\cO(\delta)\right)\right\} .\label{eq:bound6}\end{equation}
Inserting this bound in the expansion (\ref{eq:Fourier-expansion})
we get the same bound for $V^{n}u_{0}$ with an extra factor $\hbar^{-d/2}$;
however, this factor is smaller than $e^{n\delta}$ if we take $n\sim\cK\log1/\hbar$
and $\cK$ large enough, in which case the bound (\ref{eq:bound6})
also applies to $\left\Vert V^{n}u_{0}\right\Vert $. This ends the
proof of the norm bound (\ref{eq:pressure-bound}).
\end{proof}
Had we inserted (\ref{eq:hyper-1}) in the expansion (\ref{eq:Fourier-expansion})
before summing over $\balpha$, we would have obtained the following
\emph{hyperbolic dispersion estimate}:\[
\left\Vert V_{\balpha}\right\Vert \leq\min\left(b(\balpha),\, C\,\hbar^{-d/2}\, b(\balpha)\, J^{+}(\balpha)^{-1/2}\right),\qquad|\balpha|\leq\cK\log1/\hbar.\]
This type of estimate was first proved in the case of the undamped
operator $P_{0}(\hbar)$ on an Anosov manifold \cite{Anan08,AN1}.
The adaptation to the case of the damped wave equation was written
in \cite{Sche10}. A similar estimate was also shown in \cite{NZ2}
in the case of chaotic scattering.

\subsection{A {}``thickness'' condition for a spectral gap\label{sub:thickness-condition}}

In this section we present a spectral gap condition expressed only
in terms of the set of undamped trajectories (\ref{eq:undamped-set}),
but independently of the strength of the variations of the damping
(as opposed to the {}``pressure condition'' (\ref{eq:good-bound})).
A proof of this result will appear in a forthcoming publication.

To state our result in the context of a damping function $q(x,\xi)$,
we introduce the set of {}``least damped trajectories''\begin{equation}
K=\overline{\bigcup\left\{ \supp\mu,\:\mu\in\cM,\:\mu(q)=q_{+}\right\} },\label{eq:least-damped}\end{equation}
which generalizes%
\footnote{The two definitions are not strictly equal \cite{Sche11}, but this
subtlety will be irrelevant here.%
} the set (\ref{eq:undamped-set}). Our condition for a spectral gap
depends on a topological pressure $ $associated with the flow $\Phi^{t}\rest_{K}$.
We use quantities defined in $\S$\ref{sub:Anosov-reminder}.
\begin{thm}
\label{thm:thickness-bound}Let $X$ be an Anosov manifold, and $q\in S^{0}(T^{*}X)$
a damping function. Assume the following condition holds: \begin{equation}
\cP(-\varphi^{+},\Phi^{t}\rest_{K})<(d-1)\left(\frac{\nu_{\min}}{2}-\lambda_{\max}\right).\label{eq:condition11}\end{equation}
Then, there exists $\gamma>0$ ({}``the gap''), $C>0$ and $N>0$
such that, for $\hbar>0$ small enough, the following resolvent bound
holds:\begin{equation}
\forall z\in D(1/2,C\hbar)\cap\left\{ \Im z/\hbar\geq q_{+}-\gamma\right\} ,\qquad\left\Vert \left(P(\hbar)-z\right)^{-1}\right\Vert \leq C\hbar^{-N}.\label{eq:thickness-resolvent}\end{equation}
As a consequence, all eigenvalues of $P(\hbar)$ in $D(1/2,C\hbar)$
satisfy \[
\Im z_{n}/\hbar\leq q_{+}-\gamma,\]
so we have a spectral gap.
\end{thm}
In the framework of the damped wave equation with (nontrivial) damping
function $a(x)\geq0$, this gives the following 
\begin{cor}
\label{cor:thickness-resolvent}Assume that the set of undamped trajectories
(\ref{eq:undamped-set}) is such that the condition (\ref{eq:condition11})
holds. Then, there exists $\gamma>0$, $C>0$ and $N>0$ such that
\[
\left\Vert (\tau-\cA)^{-1}\right\Vert _{L^{2}\to L^{2}}\leq C\,\tau^{N},\qquad\forall\tau\in\left\{ |\Re\tau|\geq C,\:\Im\tau\geq-\gamma\right\} .\]
As a consequence, the energy decays exponentially for regular initial
data, as in (\ref{eq:decay-loss}).
\end{cor}
On a manifold of constant negative curvature $-\Lambda^{2}$, the
condition (\ref{eq:condition11}) takes the form\begin{equation}
h_{top}(\Phi^{t}\rest_{K})<(d-1)\Lambda/2,\label{eq:condition11-constcurv}\end{equation}
where $h_{top}$ is the topological entropy of the flow on $K$. It
is usually interpreted as a measure of the {}``complexity'' of the
flow on $K$; we rather see it as a measure of the {}``thickness''
of $K$. This bound reminds us of the lower bound obtained by Anantharaman
when describing the localization of eigenstates of the Laplacian on
Anosov manifolds \cite{Anan08}: in constant curvature, she shows
the semiclassical measures associated with any sequence of eigenstates
cannot be supported on a set of entropy smaller than $(d-1)\Lambda/2$,
thus forbidding the eigenstates from being too localized. Following
the thread of Remark \ref{rem:QE-link}, the spectral gap condition
(\ref{eq:condition11-constcurv}) could be seen as a {}``nonperturbative
analogue'' of this delocalization result.
\begin{rem}
For certain Anosov manifolds of dimension $d\geq3$, the condition
(\ref{eq:condition11}) can never be satisfied. Indeed, the variational
formula (\ref{eq:pressure}) shows that, for any closed invariant
set $K'\subset S^{*}X$, the pressure \[
\cP(-\varphi^{+},\Phi^{t}\rest_{K'})\geq-\sup_{\mu\in\cM}\mu(\varphi^{+})\geq-(d-1)\lambda_{\max}.\]
If the instability along $E^{+}$ is \emph{very anistropic} (meaning
that the largest and smallest positive Lyapunov exponents are very
different), then $(d-1)\lambda_{\max}-\sup_{\mu\in\cM}\mu(\varphi^{+})$
can be strictly larger than $\frac{(d-1)\nu_{\min}}{2}$, in which
case \[
\cP(-\varphi^{+},\Phi^{t}\rest_{K'})>(d-1)\left(\frac{\nu_{\min}}{2}-\lambda_{\max}\right)\qquad\mbox{for any closed invariant set }\: K'.\]

\end{rem}
This remark hints at the fact that the condition (\ref{eq:condition11})
is probably not sharp (even in 2 dimensions), except maybe on manifolds
of constant curvature. Rivière has recently improved Anantharaman's
lower bound on the support of semiclassical measures in variable curvature
\cite{Riv11}: he shows that any such support $S$ must satisfy $\cP(-\varphi^{+}/2,\Phi^{t}\rest_{S})\geq0$.
If we follow the above analogy (and also following the results of
$\S$\ref{sub:Schenck-stab}), it seems natural to expect the following
\begin{conjecture}
If the set $K$ of least damped trajectories satisfies \begin{equation}
\cP(-\varphi^{+}/2,\Phi^{t}\rest_{K})<0,\label{eq:gap-conjecture}\end{equation}
then there is a spectral gap in the spectrum of $P(\hbar)$, in the
sense of Thm \ref{thm:thickness-bound}.
\end{conjecture}
In constant curvature the condition (\ref{eq:gap-conjecture}) is
equivalent with (\ref{eq:condition11-constcurv}). In variable curvature,
it is weaker than (\ref{eq:condition11}). A Proof of (\ref{eq:gap-conjecture})
should make use of \emph{local} expansion rates, instead of the globally
defined rates $\lambda_{\max}$ and $\nu_{\min}$, like in Rivière's
work on 2-dimensional Anosov manifolds \cite{Riv10}.

In the next subsection we sketch the proof of Thm \ref{thm:thickness-bound}.

\subsubsection{Sketch of proof for Thm \ref{thm:thickness-bound}}

As was the case for Thm \ref{thm:pressure-resolvent}, the proof proceeds
by bounding the norm of the propagator $V^{t}\chi_{0}(P_{0})$ for
some logarithmic time (now the double of the Ehrenfest time, see (\ref{eq:norm-bound4})).
In the course of the proof we will also need to control the evolution
of a certain type of {}``microscopic'' Lagrangian states. The novelty
compared to the proof of Thm \ref{thm:pressure-resolvent} is that
we will now distinguish between two types of phase space points, the
{}``weakly damped'' vs. {}``strongly damped'' points.

Without loss of generality%
\footnote{Lemma \ref{lem:V^t chi_0} can be extended to logarithmic times.%
}, we may assume that the damping function $q(x,\xi)$ is compactly
supported inside $p_{0}^{-1}\left(\left[1/2\pm\eps\right]\right)$.
From now on we denote the Ehrenfest time by $T\defeq T_{Ehr}$ . We
fix some level $\alpha\in(\bar{q},q_{+})$, and consider the set of
{}``weakly damped points''\[
\Omega_{+,\alpha}\defeq\left\{ \rho\in T^{*}X,\;\la q\ra_{T,sym}(q)\geq\alpha\right\} \]
(remember that $\la q\ra_{t,sym}$ is the symmetric time average (\ref{eq:symm-average})).
The large deviation estimate (\ref{eq:large-deviation2}) provides
a bound on the volume of $\Omega_{+,\alpha}\cap S^{*}X$:\begin{equation}
\mu_{L}(\Omega_{+,\alpha}\cap S^{*}X)\leq\hbar^{-\frac{\tilde{H}(\alpha)}{\lambda_{\max}}-\cO(\eps)}.\label{eq:volume-estimate}\end{equation}
From the definition (\ref{eq:tildeH(s)}), $\tilde{H}(q_{+})$ is
equal to the pressure $\cP(-\varphi^{+},\Phi^{t}\rest_{K})$ appearing
in (\ref{eq:condition11}). If we assume (\ref{eq:condition11}),
then, by continuity of $\tilde{H}(s)$ on $\left[q_{-},q_{+}\right]$,
we may choose $\alpha\in(\bar{q},q_{+})$ large enough such that \begin{equation}
\beta(\alpha)\defeq(d-1)\left(\frac{\nu_{\min}}{2}-\lambda_{\max}\right)-\tilde{H}(\alpha)>0\,.\label{eq:condition0}\end{equation}
We will now associate a quantum projector to the set $\Omega_{+,\alpha}$.
We first symmetrize the factorization (\ref{eq:factorization}), by
writing \begin{equation}
V^{t}=U^{t/2}B_{s}(t)U^{t/2},\qquad\mbox{where}\quad B_{s}(t)\quad\mbox{has principal symbol }b_{s}(t)=e^{t\la q\ra_{t,sym}}.\label{eq:M^2n}\end{equation}
Although this symbol is positive, the operator $B_{s}(t)$ may not
be selfadjoint; we then take its polar decomposition \[
B_{s}(t)=W(t)A(t)\,,\quad\mbox{where }\begin{cases}
A(t)=\left(B_{s}(t)^{*}B_{s}(t)\right)^{1/2} & \mbox{is definite positive,}\\
\qquad W(t) & \mbox{is unitary.}\end{cases}\]
For the same reasons as in $\S$\ref{sub:Fractal-Weyl-upper} (and
due to the support assumption on $q$), the operators $B_{s}(T)$,
$A(T)$, $W(T)$ remain {}``good'' PDO up to the Ehrenfest time:
$B_{s}(T),\, A(T)\in\hbar^{-C}\Psi_{1/2-\eps}^{0}(X)$, and $W(T)\in\Psi_{1/2-\eps}^{0}(X)$. 

The operators $A(T)$ and $B_{s}(T)$ have the same leading symbol
$b_{s}(T)=e^{T\la q\ra_{T,sym}}$. Hence, to the set $\Omega_{+,\alpha}=\left\{ b_{s}(T,\rho)\geq e^{\alpha T}\right\} $
we associate the spectral projector \[
\Pi_{+}=\Pi_{+,\alpha}\defeq\bbbone_{A(T)\geq e^{\alpha T}}\,,\qquad\mbox{and call }\;\Pi_{-}=I-\Pi_{+}.\]
We use these projectors to decompose the propagator at time $2T$:
for some energy cutoff $\chi_{0}$ we write\begin{align}
V^{2T} & \chi_{0}(P_{0})=U^{T/2}W(T)\, A(T)\, U^{T}W(T)\, A(T)\, U^{T/2}\chi_{0}(P_{0})\label{eq:M^4n}\\
 & =U^{T/2}W(T)\, A(T)\left(\Pi_{+}+\Pi_{-}\right)U^{T}W(T)\left(\Pi_{+}+\Pi_{-}\right)A(T)\, U^{T/2}\chi_{0}(P_{0}).\nonumber \end{align}
The RHS splits into four terms. Three terms contain at least one factor
$\Pi_{-}$: for them we use the obvious bound \begin{equation}
\left\Vert A(T)\,\Pi_{-}\right\Vert =\left\Vert \Pi_{-}\, A(T)\right\Vert \leq e^{\alpha T}.\label{eq:Pi-}\end{equation}
The remaining term contains the factor \[
U_{++}^{T}\defeq\Pi_{+}U^{T}W(T)\,\Pi_{+}.\]
The norm of this operator will be bounded by the following \emph{hyperbolic
dispersion estimate}, the proof of which is sketched in $\S$\ref{sub:A-norm-bound}. 
\begin{prop}
\label{lem:HDE2}Assume that for some $\alpha\in(\bar{q},q_{+})$
the condition (\ref{eq:condition0}) holds. Fix some $\eps>0$. Then,
if the energy cutoff $\chi_{1}$ has small enough support, for $\hbar>0$
small enough one has \begin{equation}
\left\Vert U_{++}^{T}\,\chi_{1}(P_{0})\right\Vert _{L^{2}\to L^{2}}\leq\hbar^{\frac{\beta(\alpha)}{\lambda_{\max}}-\cO(\eps)}.\label{eq:HDE2}\end{equation}

\end{prop}
Since propagation does not modify the energy localization, if choose
in (\ref{eq:M^4n}) a cutoff $\chi_{0}\prec\chi_{1}$, we then have
\[
A(T)\, U^{T/2}\chi_{0}(P_{0})=\chi_{1}(P_{0})\, A(T)\, U^{T/2}\chi_{0}(P_{0})+\cO(\hbar^{\infty}).\]
 Inserting this identity and the bounds (\ref{eq:V^t chi-bound}),
(\ref{eq:Pi-}) and (\ref{eq:HDE2}) in the identity (\ref{eq:M^4n}),
we get \begin{equation}
\left\Vert V^{2T}\chi_{0}(P_{0})\right\Vert \leq e^{2T(q_{+}+\cO(\eps))}\left(e^{-T\beta(\alpha)}+e^{T(\alpha-q_{+})}+e^{2T(\alpha-q_{+})}\right).\label{eq:norm5}\end{equation}
We may optimize this upper bound over the level $\alpha$: the optimal
value of the exponent is reached for the (unique) parameter $\alpha_{c}\in(\bar{q},q_{+})$
solving \[
\beta(\alpha)=q_{+}-\alpha.\]
For any $\gamma>0$ satisfying $q_{+}-\gamma>\frac{q_{+}+\alpha_{c}}{2}$,
we get (for $\hbar>0$ small enough) the following norm bound for
the propagator: \begin{equation}
\left\Vert V^{2T}\chi_{0}(P_{0})\right\Vert \leq e^{2T(q_{+}-\gamma)}\,.\label{eq:norm-bound4}\end{equation}
The proof of the resolvent estimate (\ref{eq:thickness-resolvent})
is then rather straightforward.

\subsubsection{Proof of the norm bound for $U_{++}^{T}$\label{sub:A-norm-bound}}

In this last subsection we sketch the proof of Proposition \ref{lem:HDE2},
that is obtain an upper bound for \begin{equation}
\left\Vert U_{++}^{T}\,\chi_{1}(P_{0})\right\Vert _{L^{2}\to L^{2}}=\left\Vert \Pi_{+}U^{T}W(T)\Pi_{+}\chi_{1}(P_{0})\right\Vert ,\qquad T=T_{Ehr}.\label{eq:U++norm}\end{equation}
We recall that $\Pi_{+}$ is the projector associated with the region
$\Omega_{+,\alpha}$.
Let us indicate that a similar type of dispersion estimate was used by S.~Brooks,
when studying the delocalization of the eigenstates of quantized hyperbolic
automorphisms of the 2-dimensional torus (the so-called ``quantum cat maps'') \cite{Bro10}.

To estimate this norm, it will be useful to
replace this projector by a \emph{smoothed} microlocal projector obtained
by quantizing a symbol $\chi_{+}=\chi_{+,\alpha}\in S_{1/2-\eps}^{-\infty}(T^{*}X)$,
such that $\Oph(\chi_{+})$ {}``dominates'' $\Pi_{+}$:\[
\Pi_{+}\chi_{1}(P_{0})=\Oph(\chi_{+})\,\Pi_{+}\chi_{1}(P_{0})+\cO(\hbar^{\infty}).\]
The norm (\ref{eq:U++norm}) can then be bounded by:\[
\left\Vert U_{++}^{T}\chi_{1}(P_{0})\right\Vert \leq\sup_{\left\Vert u_{1}\right\Vert =\left\Vert u_{2}\right\Vert =1}\left|\la\Oph(\chi_{+})u_{2},U^{T}W(T)\Oph(\chi_{+})u_{1}\ra\right|+\cO(\hbar^{\infty})\,.\]
The symbol $\chi_{+}$ can be chosen supported inside a set of the
form $\Omega_{+,\alpha-C\eps}\cap p_{0}^{-1}([1/2\pm\eps])$. This
set is quite irregular, and the main information we have on it is
an estimate on its volume (using large deviation estimates like (\ref{eq:volume-estimate})).
It is then convenient to use an \emph{anti-Wick} quantization scheme
for $\Oph(\chi_{+})$, that is use a family of coherent states (Gaussian
wavepackets) $\left\{ e_{\rho},\:\rho\in T^{*}X\right\} $%
\footnote{Because $\chi_{+}$ is supported inside $p_{0}^{-1}(\left[1/2-\delta\right]),$it
is sufficient to construct a family of coherent states with $\rho$
in this (compact) region. %
}, to define \[
\Oph(\chi_{+})\defeq\int\frac{d\rho}{(2\pi\hbar)^{d}}\,\chi_{+}(\rho)\,\la e_{\rho},\bullet\ra\, e_{\rho}\,.\]
Each coherent state $e_{\rho}$ is normalized, and is microlocalized
in a {}``microscopic ellipse'' around $\rho$. This ellipse is chosen
to be {}``adapted'' to the flow. Let us describe it using the local
Darboux coordinates $\left\{ (y,\eta)\right\} $ as in $\S$\ref{sub:splitting-paths}.
The ellipse is {}``short'' in the energy direction, $\Delta\eta_{1}\sim\hbar^{1-\eps/2}$,
and {}``long'' in the time direction, $\Delta y_{1}\sim\hbar^{\eps/2}$.
The spread along the transverse directions is chosen isotropic: $\Delta y_{j}=\Delta\eta_{j}\sim\hbar^{1/2}$,
$j=2,\ldots,d$. 

The scalar product $\la\Op^{+}(\chi_{+})u_{2},U^{T}W_{s}(T)\Op^{+}(\chi_{+})u_{1}\ra$
can now be expressed as a double phase space integral\[
\int\int_{p_{0}^{-1}(\left[1/2\pm\eps\right])}\frac{d\rho_{1}\, d\rho_{2}}{(2\pi\hbar)^{2d}}\,\la u_{2},e_{\rho_{2}}\ra\,\la e_{\rho_{1}},u_{1}\ra\,\chi_{+}(\rho_{2})\,\chi_{+}(\rho_{1})\la U^{-T/2}\, e_{\rho_{2}},U^{T/2}W(T)\, e_{\rho_{1}}\ra.\]
The state $W(T)e_{\rho_{1}}$ is approximately identical to $e_{\rho_{1}}$.
We can then precisely describe the (undamped) evolutions of the coherent
states $e_{\rho_{i}}$ up to the times $\pm T/2$. The state $U^{T/2}e_{\rho_{1}}$
(resp. $U^{-T/2}e_{\rho_{2}}$) is a (microscopic) Lagrangian state
along a leaf of the weak unstable manifold of volume $\sim\hbar^{(d-1+\eps)/2}J_{T/2}^{+}(\rho_{1})$
centered at $\Phi^{T/2}(\rho_{1})$, (resp. a leaf of the weak stable
manifold of volume $\sim\hbar^{(d-1+\eps)/2}J_{T/2}^{+}(\Phi^{-T/2}(\rho_{2}))$
centered at $\Phi^{-T/2}(\rho_{2})$). Using the sharp energy localization
of these states and the fact that stable and unstable manifolds intersect
transversely to each other, one gets the following bound:\[
\left|\la U^{-T/2}\, e_{\rho_{2}},U^{T/2}\, W(T)\, e_{\rho_{1}}\ra\right|\leq\frac{\theta\left(\frac{\eta_{1}(\rho_{1})-\eta_{1}(\rho_{2})}{\hbar^{1-\eps}}\right)}{\,\sqrt{J_{T/2}^{+}(\rho_{1})J_{T/2}^{+}(\Phi^{-T/2}(\rho_{2}))}}+\cO(\hbar^{\infty}),\]
for some $\theta\in C_{c}^{\infty}([-1,1])$. The denominator is bounded
below by $e^{\nu_{\min}(d-1)T/2}$. Inserting this bound in the above
double integral and using large deviation estimates similar with (\ref{eq:volume-estimate})
for the support of $\chi_{+}$, one finally gets (after some manipulations)
the bound (\ref{eq:HDE2}). $\hfill\square$

\end{document}